\documentclass[aps,prl,twocolumn,showpacs,amsmath]{revtex4}
\usepackage{amsmath}
\usepackage{latexsym}
\usepackage{amssymb}
\usepackage{graphicx}
\usepackage{bm}

\newcommand{\pdag}{{\phantom{\dagger}}}
\newcommand{\bq}{\begin{equation}}
\newcommand{\eq}{\end{equation}}
\newcommand{\bn}{\begin{eqnarray}}
\newcommand{\en}{\end{eqnarray}}

\begin{document}

\title{Pumped spin-current and shot noise spectra in a single quantum dot}

\author{Bing Dong$^{1,2}$, H.L. Cui$^{1,3}$, and X.L. Lei$^{2}$} 
\affiliation{$^{1}$Department of Physics and Engineering Physics, Stevens Institute of 
Technology, Hoboken, New Jersey 07030 \\
$^{2}$Department of Physics, Shanghai Jiaotong University,
1954 Huashan Road, Shanghai 200030, China \\
$^{3}$School of Optoelectronics Information Science and Technology, Yantai University, 
Yantai, Shandong, China}

\begin{abstract}

We exploit the pumped spin-current and current noise spectra under equilibrium condition 
in a single quantum dot connected to two normal leads, as an electrical scheme for 
detection of the electron spin resonance (ESR) and decoherence. We propose spin-resolved 
quantum rate equations with correlation functions in Laplace-space for the analytical 
derivation of the zero-frequency atuo- and cross-shot noise spectra of charge- and 
spin-current. Our results show that in the strong Coulomb blockade regime, ESR-induced 
spin flip generates a finite spin-current and the quantum partition noises in the 
absence of net charge transport. Moreover, spin shot noise is closely related to the 
magnetic Rabi frequency and decoherence and would be a sensitive tool to measure them. 

\end{abstract}


\maketitle

\textit{Introduction}---There have been extensive investigations focused on electron 
spin detection and measurement via charge transport in mesoscopic quantum dot (QD) 
system,\cite{Engel,Balatsky,Martin} which is motivated from the fact that easy 
preparation and manipulation of electron spins, as well as the remarkably long spin 
coherence time, provide the way for applications in spintronics and quantum information 
processing.\cite{Prinz} In these systems, transport is governed not only by the charge 
flow, but also by the spin dynamics. Recently, a pure spin current has been reported by 
direct optical injection without generation of a net charge current.\cite{Stevens} 
Theoretically, a spin source device has also been proposed to carry pure spin flow based 
on electron spin resonance (ESR) in a QD-lead system with sizable Zeeman 
splitting.\cite{Wang1,Zhang} Moreover, both auto- and cross-correlation noise spectra of 
spin-current have been studied for this QD-based spin battery.\cite{Wang2}   

However, Coulomb blockade (CB) effect on this QD spin battery needs investigation, since 
CB effect is an essential feature of transport through a QD. More importantly, the spin 
configuration of the confined electrons on QD is apparently affected by CB effect, which 
directly determine the efficiency of this device. As well, these previous studies also 
neglected the inevitable spin decoherence due to coupling of the single spin with 
environment. Therefore, it is the purpose of this paper to study this ERS-pumped 
spin-current and its fluctuations for a QD connected with two normal leads in the strong 
CB regime at zero temperature. The setup is schematically depicted in Fig.~1: The single 
electron levels in the dot are split by an external magnetic field $B$, 
$\epsilon_{\downarrow}-\epsilon_{\uparrow}=g_{z}\mu_{B}B$ ($=\Delta$ Zeeman energy), 
where $g_{z}$ is the effective electron g-factor in the $z$ direction and $\mu_{B}$ is 
the Bohr magneton. The gate voltage controls the chemical potentials $\mu$ of two leads 
located between the two split levels and no bias voltage is applied to the two leads. 
After a spin-up electron tunnels into the QD, an oscillating magnetic field $B_{\rm 
rf}(t)=(B_{\rm rf}\cos\Omega t,\, B_{\rm rf}\sin\Omega t$) applied perpendicularly to 
the constant field $B$ with the frequency $\Omega$ nearly equal to $\Delta$ can pump 
electron to the higher level where its spin is flipped, then the spin-down electron can 
tunnel out to the leads. Because the two currents with opposite directions have 
different spin orientations, spin-currents are established in both leads with equal 
values. If the Coulomb interaction in the QD is strong enough to prohibit the double 
occupation, no more electrons can enter the QD before the spin-down electron exits. As a 
result, the number of electrons exiting from the QD is equal to that of electrons 
entering the QD, namely, the charge currents exactly cancel out each other. Otherwise, 
in the case $\epsilon_{\uparrow}+U < \mu$ the ESR pumping will generate both 
spin-current and charge-current. Note that both leads have contributions to the 
spin-current of one lead, causing the correlation of spin-currents in different leads, 
which provides possibility of studying cross-correlation of spin currents without bias 
voltage.\cite{Sauret}            

\begin{figure}[t]
\includegraphics[height=5cm,width=6cm]{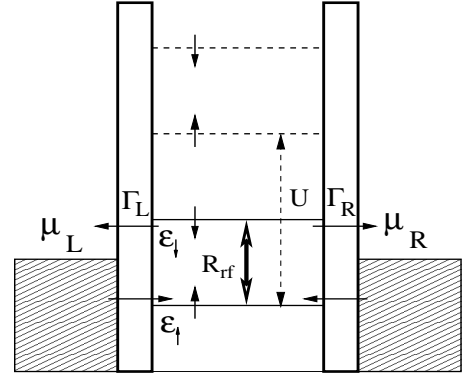}
\caption{Schematic view for the QD-based spin battery.}
\label{fig1}
\end{figure}

\textit{Theoretical formalism}---We write the Hamiltonian of the ESR-induced spin 
battery under consideration as:
\bn
H &=& \sum_{\eta, k, \sigma}\epsilon _{\eta k\sigma} 
c_{\eta k\sigma }^{\dagger }c_{\eta k\sigma }^{\pdag}+ \sum_{\sigma} \epsilon_{\sigma} 
c_{d \sigma }^{\dagger }c_{d \sigma }^{\pdag} + Un_{d \uparrow }n_{d \downarrow } \cr 
&& + \sum_{\eta, k, \sigma} (V_{\eta} c_{\eta k\sigma }^{\dagger }c_{d \sigma}^{\pdag} 
+{\rm {H.c.}}) + H_{\rm rf}(t),
\label{hamiltonian}
\en
where $c_{\eta k \sigma}^{\dagger}$ ($c_{\eta k \sigma }$) and $c_{d \sigma}^{\dagger}$ 
($c_{d \sigma}$) are the creation (annihilation) operators for electrons with momentum 
$k$, spin-$\sigma$ and energy $\epsilon_{\eta k \sigma}$ in the lead $\eta$ ($={\rm 
L,R}$) and for a spin-$\sigma$ electron on the QD, respectively. The third term 
describes the Coulomb interaction among electrons on the QD. $n_{d\sigma}=c_{d 
\sigma}^{\dagger} c_{d \sigma}^{\pdag}$ is the occupation operator in the QD. The fourth 
term represents the tunneling coupling between the QD and the reservoirs. The last term 
$H_{\rm rf}(t)$ describes the coupling between the spin states due to the rotating field 
$B_{\rm rf}(t)$ and can be written, in the rotating wave approximation (RWA), as
\bq
H_{\rm rf}(t)=R_{\rm rf} \big ( c_{d \uparrow}^{\dagger} 
c_{d\downarrow}^{\pdag}e^{i\Omega t} + c_{d\downarrow}^{\dagger} c_{d\uparrow}^{\pdag} 
e^{-i\Omega t} \big ),
\eq
with the ESR Rabi frequency $R_{\rm rf}=g_{\perp}\mu_{B}B_{\rm rf}/2$, with the g-factor 
$g_{\perp}$ and the amplitude of the rf field $B_{\rm rf}$.

Here, in order to anticipate the CB and the intrinsic spin relaxation, we utilize the 
quantum rate equations for the system density matrix elements: $\rho_{00}$ and 
$\rho_{\sigma \sigma}$ describe the occupation probability in the QD being, 
respectively, unoccupied and spin-$\sigma$ states, and the off-diagonal term 
$\rho_{\uparrow \downarrow (\downarrow \uparrow)}$ denotes coherent superposition of the 
two coupled spin states in the QD.\cite{Gurvitz} The doubly-occupied state is prohibited 
due to the infinite Coulomb interaction $U\rightarrow \infty$. Here we focus our 
interest on the nonadiabatic pumping where the photo-assisted resonance is achieved and 
neglect cotunneling processes.
For the purpose of evaluating the noise spectrum, we introduce the spin-resolved (in 
both terminals) density matrices $\rho_{ab}^{({m_{L\uparrow},m_{R\uparrow} \atop 
m_{L\downarrow},m_{R\downarrow}})}(t)$, meaning that the QD is on the electronic state 
$|a\rangle$ ($a=b=0,\,\uparrow,\,\downarrow$) or on quantum superposition state ($a\neq 
b$) at time $t$ together with $m_{L\uparrow}$ ($m_{R\uparrow}$) spin-up electrons and 
$m_{L\downarrow}$ ($m_{R\downarrow}$) spin-down electrons in the left (right) lead. 
Obviously, $\rho_{ab}(t)=\sum_{m_{L\uparrow},m_{R\uparrow} \atop 
m_{L\downarrow},m_{R\downarrow}} \rho_{ab}^{({m_{L\uparrow},m_{R\uparrow} \atop 
m_{L\downarrow},m_{R\downarrow}})}(t)$ and the quantum rate equations for these 
spin-resolved density matrices in the rotating frame with respect to $\Omega$ are
\begin{subequations}
\label{rateq}
\bn
\dot{\rho}_{00}^{({m_{L\uparrow},m_{R\uparrow} \atop m_{L\downarrow},m_{R\downarrow}})} 
& = & \Gamma_{L\downarrow} \rho_{\downarrow \downarrow}^{({m_{L\uparrow},m_{R\uparrow} 
\atop m_{L\downarrow}-1,m_{R\downarrow}})} + \Gamma_{R\downarrow} \rho_{\downarrow 
\downarrow}^{({m_{L\uparrow},m_{R\uparrow} \atop m_{L\downarrow},m_{R\downarrow}-1})} 
\cr
&& - (\Gamma_{L\uparrow}+\Gamma_{R\uparrow}) \rho_{00}^{({m_{L\uparrow},m_{R\uparrow} 
\atop m_{L\downarrow},m_{R\downarrow}})}, \label{r0} \\
\dot{\rho}_{\uparrow \uparrow}^{({m_{L\uparrow},m_{R\uparrow} \atop 
m_{L\downarrow},m_{R\downarrow}})} &=& \Gamma_{L \uparrow} 
\rho_{00}^{({m_{L\uparrow}+1,m_{R\uparrow} \atop m_{L\downarrow},m_{R\downarrow}})} + 
\Gamma_{R\uparrow} \rho_{00}^{({m_{L\uparrow},m_{R\uparrow}+1 \atop 
m_{L\downarrow},m_{R\downarrow}})} \cr
&& + \frac{1}{2T_1} \rho_{\downarrow \downarrow}^{({m_{L\uparrow},m_{R\uparrow} \atop 
m_{L\downarrow},m_{R\downarrow}})} - \frac{1}{2T_1} \rho_{\uparrow 
\uparrow}^{({m_{L\uparrow},m_{R\uparrow} \atop m_{L\downarrow},m_{R\downarrow}})} \cr
&& + i R_{\rm rf} (\rho_{\uparrow \downarrow}^{({m_{L\uparrow},m_{R\uparrow} \atop 
m_{L\downarrow},m_{R\downarrow}})} - \rho_{\downarrow 
\uparrow}^{({m_{L\uparrow},m_{R\uparrow} \atop m_{L\downarrow},m_{R\downarrow}})}), 
\label{r1} \\
\dot{\rho}_{\downarrow \downarrow}^{({m_{L\uparrow},m_{R\uparrow} \atop 
m_{L\downarrow},m_{R\downarrow}})} &=& -(\Gamma_{L\downarrow} + \Gamma_{R \downarrow}) 
\rho_{\downarrow \downarrow}^{({m_{L\uparrow},m_{R\uparrow} \atop 
m_{L\downarrow},m_{R\downarrow}})} + \frac{1}{2T_1} \rho_{\uparrow 
\uparrow}^{({m_{L\uparrow},m_{R\uparrow} \atop m_{L\downarrow},m_{R\downarrow}})} \cr 
&& \hspace{-2cm} - \frac{1}{2T_1} \rho_{\downarrow 
\downarrow}^{({m_{L\uparrow},m_{R\uparrow} \atop m_{L\downarrow},m_{R\downarrow}})} + i 
R_{\rm rf} (\rho_{\uparrow \downarrow}^{({m_{L\uparrow},m_{R\uparrow} \atop 
m_{L\downarrow},m_{R\downarrow}})} - \rho_{\downarrow 
\uparrow}^{({m_{L\uparrow},m_{R\uparrow} \atop m_{L\downarrow},m_{R\downarrow}})}), 
\label{r2} \\   
\dot{\rho}_{\uparrow \downarrow}^{({m_{L\uparrow},m_{R\uparrow} \atop 
m_{L\downarrow},m_{R\downarrow}})} &=& iR_{\rm rf} (\rho_{\uparrow 
\uparrow}^{({m_{L\uparrow},m_{R\uparrow} \atop m_{L\downarrow},m_{R\downarrow}})} - 
\rho_{\downarrow \downarrow}^{({m_{L\uparrow},m_{R\uparrow} \atop 
m_{L\downarrow},m_{R\downarrow}})}) \cr
&& \hspace{-2.5cm} + i \delta_{\rm ESR} \rho_{\uparrow 
\downarrow}^{({m_{L\uparrow},m_{R\uparrow} \atop m_{L\downarrow},m_{R\downarrow}})}
- [\frac{1}{2} ( \Gamma_{L\downarrow} + \Gamma_{R \downarrow} ) + \frac{1}{T_2}] 
\rho_{\uparrow \downarrow}^{({m_{L\uparrow},m_{R\uparrow} \atop 
m_{L\downarrow},m_{R\downarrow}})}, \label{r3} \en
\end{subequations}
and the normalization relation $\rho_{00}+ \sum_{\sigma} \rho_{\sigma \sigma}=1$. The 
equation of motion for $\rho_{\downarrow \uparrow}^{({m_{L\uparrow},m_{R\uparrow} \atop 
m_{L\downarrow},m_{R\downarrow}})}$ can be obtained by implementing complex conjugate on 
Eq.~(\ref{r3}). In these equations, $\delta_{\rm ESR}=\Delta-\Omega$ is the ESR detuning 
and $\Gamma_{\eta \sigma}=2\pi \sum_{k} |V_{\eta}|^2\delta (\omega-\epsilon_{\eta k 
\sigma})$ denotes the strength of coupling between the QD and the lead $\eta$ involving 
spin $\sigma$. In wide band limit, these tunneling amplitudes are independent of energy 
and for simplicity we set $\Gamma_{\eta \uparrow}=\Gamma_{\eta \downarrow}=\Gamma/2$. 
Furthermore, we describe the coupling of single spin with the environment in a 
phenomenological way via introducing two time scales: the spin relaxation time $T_1$ of 
an excited spin state into the thermal equilibrium, and the spin decoherence time $T_2$ 
related to the loss of phase coherence of the spin superposition state. The measurement 
of $T_2$ in QDs is currently active topic because it is the limit time scale for 
coherent spin manipulation and thus quantum information processing. In typical GaAs QD, 
$\Gamma=8~\mu$eV, corresponding to a tunneling time $\tau\sim \Gamma^{-1}=0.5$~ns. To 
observe single- and multi-photon effects in tunneling (nonadiabatic regime), it must 
require the spin flipping time $T_{\Omega}<\tau$. If one takes $B=1~$T, the Zeeman 
energy is $\Delta=26~\mu$eV which determine the optimal driving frequency 
$\Omega=13\pi~$GHz and the corresponding $T_{\Omega}=0.15~$ns. These proposed 
spin-device parameters can be easily realized by present technology.     

Recently, $T_1$ in a single QD was probed to be an order of microsecond via transport 
experiment,\cite{Fujisawa} which is notably longer than other time scales $T_1\gg 
T_2,\,\tau,\,T_{\Omega}$. Hence it is a good approximation to assume $T_1\rightarrow 
\infty$ in the following calculation. Consequently, one recovers the usual quantum rate 
equations for the reduced density matrix elements $\rho_{ab}(t)$ for a single QD with 
spin coupling: $\dot {\bm \rho}(t)=(\dot {\rho}_{00}, \dot {\rho}_{\uparrow \uparrow}, 
\dot {\rho}_{\downarrow \downarrow}, \dot {\rho}_{\uparrow \downarrow}, \dot 
{\rho}_{\downarrow \uparrow})={\cal M} {\bm \rho}(t)$ with\cite{Martin,Gurvitz}  
\bq
{\cal M}= \left (
\begin{array}{ccccc}
-\Gamma_{\uparrow} & 0 & \Gamma_{\downarrow} & 0 & 0 \\
\Gamma_{\uparrow} & 0 & 0 & iR_{\rm rf} & -iR_{\rm rf} \\
0 & 0 & -\Gamma_{\downarrow} & -iR_{\rm rf} & iR_{\rm rf} \\
0 & iR_{\rm rf} & -iR_{\rm rf} & - \Gamma_{\downarrow}V + i\delta_{\rm ESR} & 0 \\
0 & -iR_{\rm rf} & iR_{\rm rf} & 0 & - \Gamma_{\downarrow}V + i\delta_{\rm ESR}
\end{array}
\right),
\eq
and $V=\frac{1}{2} + \frac{1}{\Gamma_{\downarrow} T_2}$.

\textit{Spin-current}---The spin-related currents $I_{\eta \sigma}$ can be evaluated by 
the time change rate of spin-$\sigma$ electron number in the $\eta$ lead
\bn
I_{\eta \sigma} &=& e\dot {N}_{\eta \sigma} = e\frac{d}{dt} 
\sum_{m_{L\uparrow},m_{R\uparrow} \atop m_{L\downarrow},m_{R\downarrow}} m_{\eta \sigma} 
P(t) {\big |}_{t\rightarrow \infty}, \label{cur} \\
P(t) &=& \rho_{00}^{({m_{L\uparrow},m_{R\uparrow} \atop 
m_{L\downarrow},m_{R\downarrow}})}(t) + \rho_{\uparrow 
\uparrow}^{({m_{L\uparrow},m_{R\uparrow} \atop m_{L\downarrow},m_{R\downarrow}})}(t)
+ \rho_{\downarrow \downarrow}^{({m_{L\uparrow},m_{R\uparrow} \atop 
m_{L\downarrow},m_{R\downarrow}})}(t). \nonumber 
\en
In particular, using Eqs.~(\ref{rateq}) we find $I_{L\uparrow}=-e\Gamma_{L\uparrow} 
\rho_{00}$ and $I_{L\downarrow}= e \Gamma_{L \downarrow}\rho_{\downarrow \downarrow}$. 
The stationary solution of Eqs.~(\ref{rateq}) is:
\bn
\rho_{\uparrow \uparrow} &=& \frac{2R_{\rm rf}^2 V + \delta_{\rm ESR}^2 + \Gamma^2 
V^2}{\Xi},\\  
\rho_{\downarrow \downarrow} &=& \frac{2R_{\rm rf}^2 V} {\Xi},\\
\rho_{\uparrow \downarrow} &=& \frac{R_{\rm rf}[i \Gamma V - \delta_{\rm ESR}]}{\Xi},
\en
with $\Xi=6R_{\rm rf}^2 V + \Gamma^2 V^2 + \delta_{\rm ESR}^2$. Obviously, the 
stationary charge-current is exactly zero $I_{\eta}^{c}=I_{\eta\uparrow}+ I_{\eta 
\downarrow}=0$, while the spin-current is 
\bq
I_{L}^{s}=I_{L \downarrow} - I_{L\uparrow}= e \frac{2R_{\rm rf}^2 \Gamma 
V}{\Xi},\,\,\,\,I_{R}^{s}=-I_{L}^{s}.
\eq
Explicitly, the spin-current is proportional to the excitation power, which is 
consistent with the prototype of this ESR-based spin battery. And it's amplitude 
exhibits a saturation behavior with increasing driving field as a consequence of the 
nonlinear photon-absorption of a single spin [Fig.~2(a)]. The saturated value is 
$e\Gamma/3$ independent on the decoherence time $T_2$ and detuning $\delta_{\rm ESR}$. 
Figure 2(b) shows that the detailed dependence of the spin-current on the driving 
frequency is determined by the spin decoherence time $T_2$. In the inset of Fig.~2(a), 
we show that nonzero charge-current is also pumped by the driving field unless the 
double occupation is prohibited owing to the strong charging effect $\Gamma'=0$.\cite{U} 
This verifies our statement in the introduction.     

\begin{figure}[t]
\includegraphics[height=5cm,width=8cm]{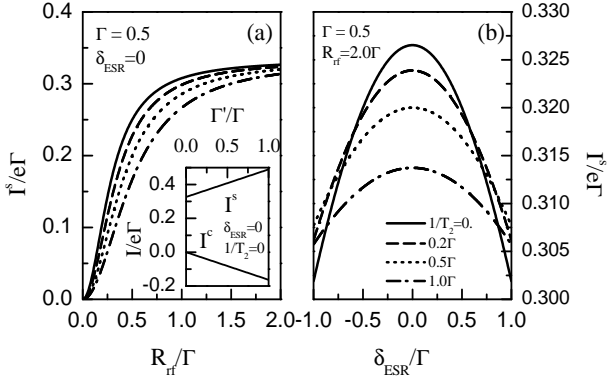}
\caption{The spin-current vs (a) strength and (b) detuning of the driving field for 
different spin decoherence time. Inset in (a): Overview of the charge- and spin-current 
with the Coulomb correlation $\Gamma'$. The calculated strength of the oscillation 
magnetic field corresponds to $0\sim 1~$T.}
\label{fig2}
\end{figure}

\textit{Spin shot noise}---Recently nonequilibrium quantum shot noise is another current 
active subject in mesoscopic physics, because the current correlation (CC) is inherently 
related to the quantization of electron charge and thus give unique information about 
electronic correlation, which cannot be obtained by probing the conductance 
only.\cite{Blanter} For instance, the cross-correlation for charge-current (the 
correlation between different terminals) is negative for a normal single-electron 
transistor, meaning anti-bunching of the wavepacket. Interestingly, much studies have 
been devoted to find a system where the cross-correlation changes its 
sign.\cite{Samuelsson} Here we would like to address that the spin-resolved CC is more 
useful to describe electron correlation, because the electronic wavepacket with opposite 
spins is uninfluenced by the Pauli exclusion principle and only reflects unambiguous 
information about the interaction. For the purpose of evaluating the charge and spin CC 
from the quantum rate equations (\ref{rateq}), we extend the MacDonald's formula for the 
spin-resolved situation:\cite{MacDonald}
\bn
S_{\eta \eta'}^{\sigma \sigma'}(\omega) &=& 2\omega e^2 \int_{0}^{\infty} dt \sin 
(\omega t) \frac{d}{dt}  {\Big \{} \sum_{m_{L\uparrow},m_{R\uparrow} \atop 
m_{L\downarrow},m_{R\downarrow}} m_{\eta \sigma} m_{\eta' \sigma'} \cr
&& \times P(t) - t^2 I_{\eta \sigma} I_{\eta' \sigma'}/e^2 {\Big \}}.
\en
Specially, the zero-frequency shot noise spectrum is
\bq
S_{\eta \eta'}^{\sigma \sigma'}(0) = 2 e^2 \frac{d}{dt}  {\Big \{} 
\sum_{m_{L\uparrow},m_{R\uparrow} \atop m_{L\downarrow},m_{R\downarrow}} m_{\eta \sigma} 
m_{\eta' \sigma'} P(t) - \frac{t^2}{e^2} I_{\eta \sigma} I_{\eta' \sigma'} {\Big 
\}}{\big |}_{t\rightarrow \infty}. \label{sn}
\eq
The auto- and cross-noise spectra of the charge-current and spin-current can be obtained 
from these correlations $S_{\eta \eta'}^{\sigma \sigma'}$: $S_{\eta \eta'}^{c/s}=S_{\eta 
\eta'}^{\uparrow \uparrow}+ S_{\eta \eta'}^{\downarrow \downarrow} \pm S_{\eta 
\eta'}^{\uparrow \downarrow} \pm S_{\eta \eta'}^{\downarrow \uparrow}$.

In order to evaluate these correlations, we introduce the following generating 
functions:
\bq
G_{\eta \sigma}^{ab}(t) = \sum_{m_{L\uparrow},m_{R\uparrow} \atop 
m_{L\downarrow},m_{R\downarrow}} m_{\eta \sigma} 
\rho_{ab}^{({m_{L\uparrow},m_{R\uparrow} \atop m_{L\downarrow},m_{R\downarrow}})}(t).
\eq
With the help of Eqs.~(\ref{rateq}), all noise spectra are therefore relevant with these 
auxiliary functions and $\rho_{ab}(t)$ as:
\bn
\frac{S_{\eta \eta'}^{\sigma \sigma'}(0)}{2e^2} &=& {\big \{} \frac{1}{2} 
\Gamma_{\sigma} \delta_{\sigma \sigma'} [\delta_{\eta \eta'} 
(\rho_{00}(t)\delta_{\sigma\uparrow} + \rho_{\downarrow \downarrow}(t) \delta_{\sigma 
\downarrow}) \cr
&& \hspace{-1cm} + \sum_{\eta''} (G_{\eta'' \sigma}^{\downarrow\downarrow}(t) 
\delta_{\sigma \downarrow} - G_{\eta'' \sigma}^{00}(t) \delta_{\sigma \uparrow})]+ 
\delta_{\sigma \bar{\sigma'}}[\frac{1}{2} \Gamma_{\downarrow} G_{\eta 
\uparrow}^{\downarrow \downarrow}(t) \cr
&& \hspace{-1cm} - \frac{1}{2} \Gamma_{\uparrow} G_{\eta' \downarrow}^{00}(t) ]  
- \frac{2t}{e^2} I_{\eta \sigma }I_{\eta' \sigma'} {\big \}} {\big |}_{t\rightarrow 
\infty}.
\en
On the other hand, using Eqs.(\ref{rateq}), the equations of motion for ${\bf G}_{\eta 
\sigma}(t)$ is explicitly obtained in matrix form: $\dot {\bf G}_{\eta \sigma}(t)= {\cal 
M} {\bf G}_{\eta \sigma}(t) + {\cal G}_{\eta\sigma} {\bm \rho}(t)$ with
\bq
{\cal G}_{\eta\sigma}= \left (
\begin{array}{ccccc}
0 & 0 & \Gamma_{\eta \downarrow}\delta_{\sigma \downarrow} & 0 & 0 \\
-\Gamma_{\eta \uparrow}\delta_{\sigma \uparrow} & 0 & 0 & 0 & 0 \\
0 & 0 & 0 & 0 & 0 \\
0 & 0 & 0 & 0 & 0 \\
0 & 0 & 0 & 0 & 0
\end{array}
\right).
\eq
Applying Laplace transform to these equations yields
\bq
{\bf G}_{\eta \sigma}(s) = (s {\bf I}-{\cal M})^{-1} {\cal G}_{\eta\sigma} {\bm 
\rho}(s),
\eq
where ${\bm \rho}(s)$ is readily obtained by performing Laplace transform on its 
equations of motion and the normalization relation. Due to inherent long-time stability 
of the physics system under investigation, all real parts of nonzero poles of ${\bm 
\rho}(s)$ and ${\bf G}(s)$ are negative definite. Consequently, the large-$t$ behavior 
of the auxiliary functions is entirely determined by the divergent terms of the partial 
fraction expansions of ${\bm \rho}(s)$ and ${\bf G}_{\eta \sigma}(s)$ at $s\rightarrow 
0$. 

Resolving both ${\bf G}_{\eta \sigma}(s)$ and ${\bm \rho}(s)$ into the partial fraction 
expansion forms and performing inverse Laplace transform, we can eventually derive the 
analytical expressions for these large-$t$ asymptotic correlations.
Surprisingly, we find the zero-frequency charge shot noises being proportional to the 
spin-current $S_{LL}^{c} = -S_{LR}^{c}= e^2 |I_{L}^{s}|$, which manifests that pumping 
processes do generate shot noise even in the absence of net charge-current. This feature 
is underlying analog of the quantum partition noise of photo-excited electron-hole pairs 
demonstrated in the recent experiment.\cite{Reydellet} Spin-up and spin-down electrons 
in this spin battery are dissociated by tunneling events in the presence of pumping 
field but without bias: spin-up electrons inject into the QD, but spin-down electrons 
flow off in an opposite direction. As expected, the cross-correlation of the charge 
noise is always negative definite, showing anti-bunching statistics, but it has the same 
magnitude with the auto-correlation of the charge-current originated from the 
conservation law of charge.\cite{Wang2}

\begin{figure}[t]
\includegraphics[height=5cm,width=8cm]{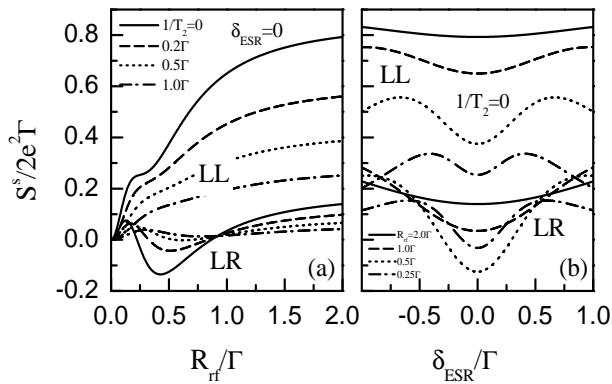}
\caption{The zero-frequency auto- and cross-correlations of spin-current vs (a) strength 
and (b) detuning of driving field.}
\label{fig3}
\end{figure}

However, the situation is very different for the zero-frequency spin shot noise, as 
shown in Fig.~3. The auto-correlation of the spin-current is surely positive definite, 
while the cross-correlation is either positive or negative depending on a number of 
parameters: the pumping amplitude $R_{\rm rf}$, the detuning $\delta_{\rm ESR}$, and the 
decoherence $T_2$. For small decoherence rates, the spin cross-correlation experiences 
the change of sign with increasing excitation $R_{\rm rf}$ at resonant pumping 
$\delta_{\rm ESR}=0$. Interestingly, this cross-correlation is always positive (1) at 
large excitation $R_{\rm rf}> \Gamma$ regardless of decoherence $T_2$ [Fig.~3(a)] or (2) 
if apart away from quantum resonance [Fig.~3(b)]. The significant role of decoherence is 
to evidently reduce these noise spectra and even to eliminate the change of sign in the 
spin cross-correlation. It is addressed that spin shot noise is more sensitive to the 
spin decoherence than the spin-current and the charge shot noise. This could provide a 
way to measure the spin decoherence time $T_2$.

In summary, we have presented a generic method for analytical calculation of the atuo- 
and cross-shot noise spectra of charge- and spin-current in an ESR-based single QD 
system. We found that in the strong CB regime, indeed ESR pumping generates a finite 
spin-current and quantum partition noises in the absence of net charge transport. And 
the spin shot noises display complicated behaviors depending on the pumping parameters 
and the spin decoherence time. The measurement of spin-current and shot noise provides a 
scheme for the Rabi oscillation and spin decoherence detections in the QD system.                 

BD and HLC are supported by the DURINT Program administered by the US Army Research 
Office. XLL is supported by Major Projects of National Natural Science Foundation of 
China, the Special Founds for Major State Basic Research Project and the Shanghai 
Municipal Commission of Science and Technology.


\begin{thebibliography}{99}
\bibitem{U}{Extension of zero-temperature spin-resolved quantum rate equations and 
current formula are easily achieved for considering the doubly occupied state. See 
Ref.~\onlinecite{Gurvitz}. Charging effect is modeled by tunneling rate $\Gamma'$.}
\end{thebibliography}
\end{document}